\documentclass[12pt]{article}
\usepackage{latexsym,amsmath,amssymb,epsfig,graphicx}
\newcommand{\be}{\begin{equation}}
\newcommand{\ee}{\end{equation}}
\newcommand{\bea}{\begin{eqnarray}}
\newcommand{\eea}{\end{eqnarray}}
\newcommand{\kma}{\; ,}
\newcommand{\pkt}{\; .}

\newcommand{\calm}{{\cal M}}

\begin{document}
\begin{titlepage}
\begin{flushright}
hep-th/yymmnn \\
December 2010
\end{flushright}
\vspace{8mm}
\begin{center}
{\Large \bf
One-loop corrections to the string tension of the vortex
in the Abelian Higgs model: Erratum.}
\\\vspace{8mm}
{\large  J\"urgen Baacke\footnote{e-mail:~
juergen.baacke@tu-dortmund.de}} \\
{  Fakult\"at Physik, Technische Universit\"at Dortmund \\
D - 44221 Dortmund, Germany
}\\
\vspace{4mm}
and\\
\vspace{4mm}
{\large  Nina Kevlishvili\footnote{e-mail:~
kevlishvili@theorie.physik.uni-goettingen.de}} \\
{Institut f\"ur theoretische Physik,
Georg-August-Universit\"at G\"ottingen,
Friedrich-Hund-Platz 1, 37077 G\"ottingen, Germany
}\\
\vspace{5mm}
%\newpage

{\bf Abstract}
\end{center}
We correct two errors in our previous computation of one-loop
corrections to the vortex string tension:
(i) the contribution of the longitudinal and timelike modes of the 
gauge fields were forgotten and are included now; (ii) a trivial error in the
numerical code has led to considerable errors in the 
subtracted integrals. We here present the corrected results. 
\end{titlepage}

%*******************************************************Introduction

While working on a related subject, one of us (J.B.) became aware
of two mistakes in our previous publication \cite{Baacke:2008sq}.\\\\
(i) The fluctuations of the longitudinal and timelike modes of the gauge 
fields have escaped our attention in the final write-up.
Their fluctuation operator is in fact
identical to the one for the Faddeev-Popov ghosts,
\be
\calm = \Box+ g^2 \phi^2
\kma
\ee 
see Eq. (3.17) of Ref. \cite{Baacke:2008sq}\footnote{The first $+$ sign on the
right hand side of that equation should be a $-$ sign.}.
The contributions of these two modes therefore cancel exactly against 
those of the Faddeev-Popov
ghosts. In our publication we have computed the correction to the
string tension of the coupled system of transverse gauge field and
Higgs field fluctuations and from these we have subtracted the
corrections due to the Faddeev-Popov ghosts. The mistake is easy to correct: 
instead of adding the forgotten timelike and transverse modes we just 
do not subtract the Faddeev-Popov ghosts any more. So the correct result 
is equal to the full contribution of the coupled system of
Higgs and transverse gauge fields. This correction has to be applied to
the renormalized leading orders $\Delta \sigma^{(1)}_{\rm fin}$
and $\Delta\sigma^{(2)}_{\rm fin}$,
as well as to the sum of the higher
order terms  $\Delta\sigma_{\rm sub}$,
which is computed numerically. 

For the first and second order diagrams this means that
Eq. (7.6) of Ref.\cite{Baacke:2008sq}
becomes
\be
\Delta \sigma^{(1)}_{\rm fin}=
-\frac{m_W^2}{32\pi^2}\left\{ (1-\ln\frac{m_W^2}{\mu^2})(3+2\xi^2)
 -\frac{3}{2}\xi^2\ln\xi^2\right\}\int d^2x (f^2(r)-1) 
\kma
\ee  
and Eq. (7.7) is changed into
\bea\nonumber
\Delta\sigma^{(2)}_{\rm fin}&=&\Delta\sigma_{\rm fl,11,11}
+\Delta\sigma_{\rm fl,22,22}+\Delta\sigma_{\rm fl,33,33}+
\Delta\sigma_{\rm fl,44,44}
\\
&&+\Delta\sigma_{\rm fl,(12)3,(12)3}+
\Delta\sigma_{\rm fl,(12)4,(12)4}+
\Delta\sigma_{\rm fl,34,34}
\pkt\eea

(ii) In the computation of the subtracted integrals which represent
the graphs of third and higher order we found a trivial mistake
in the numerical code. In the final integral  
\be
\Delta \sigma_{\rm sub}=-\int_0^\infty\frac{dp  p^3}{4\pi}F_{\rm sub}(p)
\ee 
the weight factor was $p$ instead of $p^3$ in the first 
integration interval. Unfortunately, due to the translation mode pole 
the contribution of this single interval became much larger than 
all the remaining integral which was computed correctly. This explains 
the large corrections in $\Delta\sigma_{\rm sub}$.

All in all this implies, unfortunately, that all the entries in table I 
of Ref. \cite{Baacke:2008sq} get changed,  except for the classical 
string tensions, and Fig. 20 is modified accordingly.
The corrected table and figure are presented here in Table 1 and Fig. 1.   

\begin{figure}[htb]
\vspace{15mm}
\begin{center}
\includegraphics[scale=0.5]{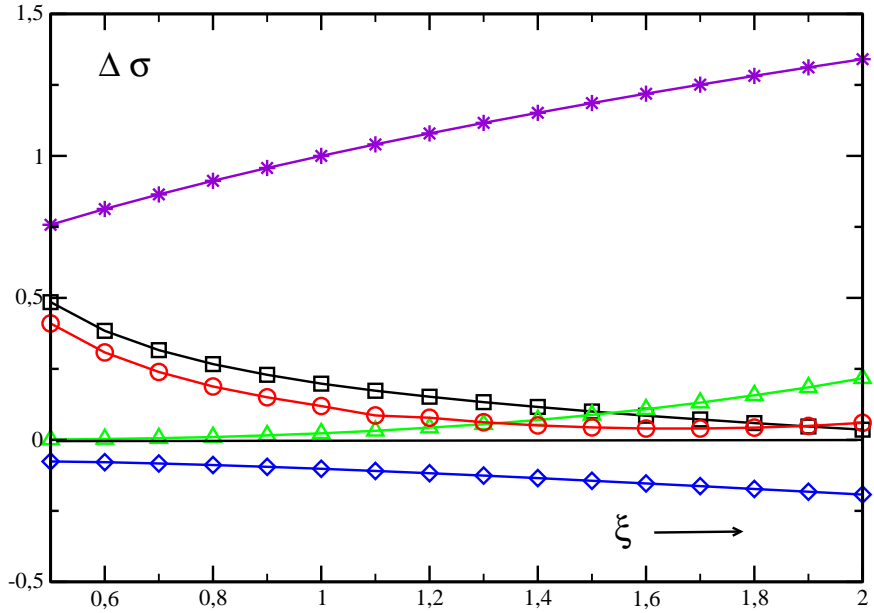}
\end{center}\vspace{9mm}
\caption{\label{fig:deltasigma}
The correction to the string tension as a function of
$\xi=m_H/m_W$. Squares: contributions with one vertex,
$\Delta \sigma^{(1)}_{\rm fin}$; triangles: contributions with two vertices
$\Delta\sigma^{(2)}_{\rm fin}$; diamonds: subtracted contribution
$\Delta\sigma_{\rm sub}$, circles: total one-loop correction
$\Delta \sigma_{\rm tot}$, asterisks: the classical string tension
mutiplied by $g^2/\pi$. All string tensions are in units of $m_W^2$.}  
\end{figure}
\begin{table}
\begin{center}
\begin{tabular}{|r|r|r|r|r|r|}
\hline
$\xi$& $\Delta \sigma^{(1)}_{\rm fin}$&$\Delta \sigma^{(2)}_{\rm fin}$&
$\Delta\sigma_{\rm sub}$&$\Delta\sigma_{\rm tot}$&$g^2\sigma_{cl}/\pi$
\\
\hline
$0.5$&.485&.001&  -.0760& .41&.75742\\
$0.6$&.384&.003&  -.0786&.308&.81306\\
$0.7$&.316&.006&  -.0832&.239&.86441\\
$0.8$&.267&.010&  -.0886&.188&.91232\\
$0.9$&.229&.016&  -.0949&.150&.95737\\
$1.0$&.198&.023&  -.1016&.119&1.0000\\
$1.1$&.173&.032&  -.1093&.086&1.0405\\
$1.2$&.152&.043&  -.1173&.078&1.0792\\
$1.3$&.133&.055&  -.1258&.062&1.1163\\
$1.4$&.116&.070&  -.1347&.051&1.1518\\
$1.5$&.100&.088&  -.1439&.044&1.1860\\
$1.6$&.085&.108&  -.1533&.040&1.2190\\
$1.7$&.072&.131&  -.1630&.040&1.2509\\
$1.8$&.059&.157&  -.1728&.043&1.2818\\
$1.9$&.047&.185&  -.1827&.049&1.3116\\
$2.0$&.036&.217&  -.1927&.060&1.3406\\
\hline
\end{tabular}
\end{center}
\vspace*{3mm}
\caption{\label{table:finalresults} The correction to the string tension as a function of
$\xi=m_H/m_W$. We present the finite part of the
 contributions with one vertex,
$\Delta \sigma^{(1)}_{\rm fin}$, the finite contributions from
graphs with two vertices $\Delta\sigma^{(2)}_{\rm fin}$, 
the sum of higher order terms
$\Delta\sigma_{\rm sub}$, and the total one-loop correction
$\Delta \sigma_{\rm tot}$. We also include the classical string tensions.
All entries are in units of $m_W^2$.} 
\end{table}

%******************************************************** References

\bibliography{novqc}

\end{document}